\documentclass[12pt]{iopart}

\expandafter\let\csname equation*\endcsname\relax
\expandafter\let\csname endequation*\endcsname\relax

\usepackage{amsmath, amssymb, graphicx, subfigure, bm, color}
\usepackage[normalem]{ulem}
\usepackage{color, hyperref}
\usepackage[all]{hypcap}

\definecolor{MyDarkGreen}{rgb}{0,0.6,0}
\definecolor{MyDarkBlue}{rgb}{0,0,0.8}
\definecolor{MyDarkRed}{rgb}{0.6,0,0.3}
\hypersetup{breaklinks=true,colorlinks=true,plainpages=true,linktocpage=true,linkcolor=MyDarkBlue,citecolor=MyDarkGreen,urlcolor=MyDarkRed,pdfborder={0 0 0},%
pdfauthor={Sang-Kil Son},%
pdftitle={Compton spectra of atoms at high x-ray intensity}%
}

\newlength{\figurewidth}
\newcommand{\degree}{\ensuremath{^\circ}}

\begin{document}

\title[Compton spectra of atoms at high x-ray intensity]{Compton spectra of atoms at high x-ray intensity}

\author{Sang-Kil Son$^{1,2}$, Otfried Geffert$^{1}$, and Robin Santra$^{1,2,3}$}

\address{$^1$ Center for Free-Electron Laser Science, DESY, Notkestrasse 85, 22607 Hamburg, Germany}
\address{$^2$ The Hamburg Centre for Ultrafast Imaging, Luruper Chaussee 149, 22761 Hamburg, Germany}
\address{$^3$ Department of Physics, University of Hamburg, Jungiusstrasse 9, 20355 Hamburg, Germany}
\ead{sangkil.son@cfel.de}
\begin{abstract}
Compton scattering is the nonresonant inelastic scattering of an x-ray photon by an electron and has been used to probe the electron momentum distribution in gas-phase and condensed-matter samples.
In the low x-ray intensity regime, Compton scattering from atoms dominantly comes from bound electrons in neutral atoms, neglecting contributions from bound electrons in ions and free (ionized) electrons.
In contrast, in the high x-ray intensity regime, the sample experiences severe ionization via x-ray multiphoton multiple ionization dynamics.
Thus, it becomes necessary to take into account all the contributions to the Compton scattering signal when atoms are exposed to high-intensity x-ray pulses provided by x-ray free-electron lasers (XFELs).
In this paper, we investigate the Compton spectra of atoms at high x-ray intensity, using an extension of the integrated x-ray atomic physics toolkit, \textsc{xatom}.
As the x-ray fluence increases, there is a significant contribution from ionized electrons to the Compton spectra, which gives rise to strong deviations from the Compton spectra of neutral atoms.
The present study provides not only understanding of the fundamental XFEL--matter interaction but also crucial information for single-particle imaging experiments, where Compton scattering is no longer negligible.
\end{abstract}


\submitto{\jpb}
\maketitle

\section{Introduction}\label{sec:introduction}

Compton scattering is the nonresonant inelastic scattering of an x-ray photon by an electron~\cite{Compton23,Compton61}.
It has been used to probe the electron momentum distribution of various samples, which is one of the fundamental quantities of interest and has a broad range of applications~\cite{Cooper85}.
With the help of x-ray synchrotron light sources, it has become feasible to accurately measure Compton scattering for atoms and molecules (for recent examples, see \cite{Sakurai11,Kobayashi11,Zhao15}).
With recent advances in synchrotron radiation facilities, the high-resolution x-ray Compton scattering technique allows us to make cutting-edge studies including visualization of bonding in liquids~\cite{Okada15} and imaging the hole state of dopants in complex materials~\cite{Sakurai11a}.

X-ray free-electron lasers (XFELs)~\cite{McNeil10,Pellegrini12,Ribic12} open up new opportunities even beyond what synchrotron light sources can offer.
The brightness of XFELs is many orders of magnitude higher than that of synchrotron sources~\cite{Schneider10}.
Unprecedentedly ultraintense x-ray pulses enable us to study nonlinear x-ray physics~\cite{Young10,Doumy11,Glover12,Tamasaku14}, including nonlinear two-photon x-ray Compton scattering~\cite{Fuchs15,Hopersky15}.
The features of ultraintense and ultrashort XFEL pulses are useful to create warm dense matter~\cite{Vinko12}, and the inelastic x-ray scattering technique has been developed to measure characteristics (density and temperature) of high energy density plasma~\cite{Glenzer03,Gregori03,Sahoo08,Glenzer09}.
One of the most prominent applications of XFELs is x-ray imaging of biological macromolecules at atomic resolution~\cite{Gaffney07,Patterson14,Schlichting15}.
The intense XFEL pulses provide enough signal to reconstruct molecular structures from nano-sized crystals and even from non-crystalline single particles~\cite{Neutze00,Chapman06,Chapman11,Seibert11}.
In crystallography, elastic x-ray scattering signals from crystals coherently interfere, thus yielding Bragg peaks, while Compton scattering gives rise to a background signal.
In single-particle imaging experiments, however, Compton scattering signals are not distinguishable from the elastic x-ray scattering signals, unless the photon detector can resolve Compton energy shifts.
The importance of the Compton background has been pointed out in the x-ray molecular imaging community~\cite{Jurek12,Ziaja12a,Slowik14,Gorobtsov15,Ziaja15}.

To advance XFEL-driven science, it is critical to understand fundamental interactions between atoms and intense x-ray pulses.
The XFEL--atom interaction is described by x-ray multiphoton ionization dynamics~\cite{Santra14}, where the atomic system absorbs many x-ray photons sequentially and ejects many electrons.
To investigate this multiphoton multiple ionization dynamics of atoms, the \textsc{xatom} toolkit has been developed~\cite{xatom}.
\textsc{xatom} calculates electronic structure based on the Hartree-Fock-Slater method for any given element and any given electronic configuration.
Further, it calculates cross sections and rates of many x-ray-induced processes including photoionization, fluorescence, and Auger (Coster-Kronig) decay for every single configuration that may be formed during and after intense x-ray pulses.
These calculated rates and cross sections serve as input data to a set of coupled rate equations that must be solved to simulate ionization dynamics.
After solving the rate equations, \textsc{xatom} generates ion spectra, electron spectra, and photon spectra~\cite{Son12f}, which may be directly compared with experiments.
Since for heavy atoms typically a huge set of rate equations are involved~\cite{Rudek12,Fukuzawa13}, \textsc{xatom} utilizes a Monte Carlo scheme~\cite{Son12f}.

\textsc{xatom} has been a key development for XFEL-driven science.
Firstly, \textsc{xatom} has served as an essential toolkit for various XFEL experiments on gas-phase atoms~\cite{Doumy11,Rudek12,Fukuzawa13,Rudek13,Motomura13} conducted at LCLS~\cite{Emma10} and at SACLA~\cite{Ishikawa12a}.
It has provided capabilities to interpret experimental data, to examine nonlinear two-photon x-ray ionization~\cite{Doumy11,Sytcheva12a}, and to propose a novel x-ray ionization mechanism involving resonant excitations and accompanying decays~\cite{Rudek12,Rudek13}.
The noble gas results obtained by \textsc{xatom} have been used to calibrate x-ray beam profiles~\cite{Rudek12,Fukuzawa13,Murphy14,Tachibana15}.
Secondly, \textsc{xatom} provides valuable information on the dynamical behavior of individual atoms within molecules using an independent atomic model.
The x-ray multiphoton ionization model has been applied to study the impact of frustrated absorption on elastic x-ray scattering dynamics~\cite{Son11a}.
The dynamical information of multiple ionization has been employed to interpret an x-ray-induced fragmentation experiment on small molecules at low x-ray intensity~\cite{Dunford12a}.
The time-dependent atomic form factors obtained from \textsc{xatom} have been used to construct diffraction patterns of nanocrystals exposed to intense x-ray pulses~\cite{Abdullah16}.
Based on knowledge of the dynamical behavior of heavy atoms, a high-intensity version of the multiwavelength anomalous diffraction method has been proposed~\cite{Son11e,Son13} to address the phase problem in femtosecond nanocrystallography with XFELs.
Furthermore various high-intensity phasing methods utilizing the selective and extensive ionization of heavy atoms have been developed~\cite{Galli15,Galli15a,Galli15b}.
Thirdly, \textsc{xatom} has been combined with other tools for modeling XFEL-induced dynamics of matter: \textsc{xmolecule}~\cite{xmolecule} for molecules and \textsc{xmdyn}~\cite{xmdyn,Jurek16} for complex systems.
The atomic electronic structure and atomic data calculated by \textsc{xatom} have served as inputs to \textsc{xmdyn} to simulate complex systems such as C$_{60}$~\cite{Murphy14} and Ar and Xe clusters~\cite{Tachibana15} interacting with intense XFEL pulses.
They have also been used for \textsc{xmolecule} to calculate molecular electronic structure~\cite{Hao15} and molecular data~\cite{Inhester16} to investigate molecular x-ray multiphoton ionization dynamics~\cite{Inhester16} and ultrafast explosion dynamics of small polyatomic molecules induced by intense XFEL pulses~\cite{Rudenko16}.
Lastly, \textsc{xatom} has been extended to treat atoms and ions immersed in a plasma environment to investigate ionization potential lowering in warm dense matter~\cite{Thiele12,Son14c}, which has further been employed for studying x-ray resonant magnetic scattering in materials~\cite{Mueller13}.
Also note that the atomic electronic continuum states that are accurately calculated by \textsc{xatom} have been used for modeling high harmonic generation of rare gas atoms~\cite{Bhardwaj13}.
Inclusion of resonant photoexcitation and of relativistic effects for heavy atoms is in progress~\cite{Toyota16}.

To investigate x-ray scattering dynamics at high x-ray intensity including severe ionization of individual atoms, \textsc{xatom} calculates elastic x-ray scattering cross sections (atomic form factors)~\cite{Son11a} and inelastic (Compton) x-ray scattering cross sections~\cite{Slowik14} of atoms.
For the latter, the cross section is given as a doubly differential expression per angle and per energy of the scattered photon.
This cross section as a function of the scattered photon energy (at a given angle of the scattered photon momentum) represents the Compton spectra.
To the best of our knowledge, Compton spectra induced by intense x-ray pulses have not been studied so far.
Challenges are a) to take into account electrons produced through massive ionization of a single atom and b) to include all possible bound-electron configurations generated from multiphoton multiple ionization dynamics.
Since the Compton scattering cross section calculation for each configuration involves enormously many recurring evaluations of spherical Bessel functions, an efficient numerical procedure is required to make the Compton spectra calculation feasible.
In this paper, we present Compton spectra of atoms exposed to intense x-ray pulses, calculated by using \textsc{xatom} equipped with a suitable numerical procedure. 

The paper is organized as follows.
Section~\ref{sec:radiation_damage} describes how to calculate Compton spectra including ionization dynamics.
In Secs.~\ref{sec:bound_electron} and \ref{sec:free_electron}, we provide expressions for the doubly differential Compton scattering cross section for bound electrons and for free electrons.
In Sec.~\ref{sec:Bessel}, we present an efficient way to calculate the spherical Bessel functions needed for Compton spectra calculations.
Section~\ref{sec:results} shows the Compton spectra of Ar and Xe, and we conclude with a summary in Sec.~\ref{sec:conclusion}.

\section{Methodology}\label{sec:methodology}

\subsection{Ionization dynamics and Compton spectra at high x-ray intensity}\label{sec:radiation_damage}

To simulate ionization dynamics, we employ the rate-equation approach~\cite{Young10,Son11a,Rohringer07,Makris09}.
This approach has been successfully applied to describing x-ray multiphoton ionization dynamics of atoms and molecules.
The transitions between electronic configurations $\lbrace I \rbrace$ are described by coupled rate equations,
\begin{equation}\label{eq:rate_equations}
\frac{d}{dt} P_I(t) = \sum_{I' \neq I}^\text{all config.} \left[ \Gamma_{I' \rightarrow I} P_{I'}(t) - \Gamma_{I \rightarrow I'} P_{I}(t) \right],
\end{equation}
where $P_I(t)$ is the time-dependent population of a given bound-electron configuration $I$ and $\Gamma_{I \rightarrow I'}$ is the transition rate from $I$ to $I'$.
All x-ray-induced transition rates and cross sections are calculated by \textsc{xatom}~\cite{xatom}.
To construct a set of rate equations, we include photoionization processes and accompanying relaxation processes such as fluorescence and Auger (Coster-Kronig) decay.
In the regime considered here, the Compton scattering cross section is one order of magnitude smaller than the photoionization cross section~\cite{Thompson01}.
Therefore, ionization via Compton scattering is not included in the set of rate equations.

When an electron is ionized by photoabsorption or Auger (Coster-Kronig) decay, its kinetic energy is assigned to an energy bin $E$.
The time-dependent population of the energy bin, $P_E(t)$, is given by
\begin{equation}\label{eq:P_E}
P_E(t) = \int_{-\infty}^{t} dt' \; \sum_{I, I'}^{E} \Gamma_{I' \rightarrow I} P_{I'}(t'),
\end{equation}
where the double sum chooses only pairs of $I$ and $I'$ where the $I' \rightarrow I$ process ejects an electron with a kinetic energy in the energy bin considered.
When $t \rightarrow \infty$, $P_E(t)$ represents an electron spectrum, assuming that ejected electrons do not interact with each other.

The Compton scattering signal $I_C$ is the pulse-weighted integral of the time-dependent Compton scattering cross section from bound and free electrons,
\begin{equation}\label{eq:scattering_signal}
I_C(\Omega_F,\omega_F)
 = \int_{-\infty}^{\infty} \! dt \, \mathcal{F} g(t) \left[ \frac{ d^2 \sigma_\text{bound} }{ d \Omega_F d\omega_F } ( t ) + \frac{ d^2 \sigma_\text{free} }{ d \Omega_F d \omega_F } ( t ) \right],
\end{equation}
where $\Omega_F$ is the solid angle of the scattered photon momentum with respect to the incident photon momentum and $\omega_F$ is the scattered photon energy.
$\mathcal{F}$ is the x-ray fluence and $g(t)$ is the normalized pulse envelope.
Then the flux is given by $\mathcal{F} g(t)$.
It is assumed that both bound and free electrons experience the same local fluence within a micron-size x-ray beam, because a free electron, for instance, with a kinetic energy of 2~keV can travel only $\sim$30~nm per fs.
The spatial distribution of the x-ray beam can be treated via integration over the interaction volume~\cite{Young10,Rudek12}.
The time-dependent doubly differential scattering cross section from bound electrons is given by taking into account all electronic configurations at the given time $t$~\cite{Slowik14},
\begin{equation}\label{eq:sigma_config}
\frac{d^2 \sigma_\text{bound}}{ d \Omega_F d \omega_F } ( t ) = \sum_I^\text{all config.} P_I(t) \frac{ d^2 \sigma_{\text{bound},I} }{ d \Omega_F d \omega_F },
\end{equation}
where $P_I(t)$ is obtained from solving Eq.~\eref{eq:rate_equations}, and $\sigma_{\text{bound},I}$ is the Compton scattering cross section for the given configuration $I$. 
As discussed later, the doubly differential Compton scattering cross section calculations for all electronic configurations are formidably expensive.
Therefore, the time-dependent cross section is approximated using the charge states $\lbrace q \rbrace$ (rather than the full configuration information $\lbrace I \rbrace$), 
\begin{equation}\label{eq:sigma_charge}
\frac{d^2 \sigma_\text{bound}}{ d \Omega_F d \omega_F } ( t ) \approx \sum_q^\text{all charges} P_q(t) \frac{ d^2 \sigma_{\text{bound},q} }{ d \Omega_F d \omega_F },
\end{equation}
where $P_q(t) = \sum_I^{\text{charge($I$)} = q} P_I(t)$ and $\sigma_{\text{bound},q}$ is the Compton scattering cross section for the ground-state configuration for the given charge $q$.
A similar approach was applied to elastic x-ray scattering~\cite{Galli15b}.
The summation over all the charge states is much simpler than that over all the configurations.
The doubly differential scattering cross section $\sigma_{\text{bound}}$ will be given in Sec.~\ref{sec:bound_electron} [Eq.~\eref{eq:DDSCS_full_angle}].
A numerical comparison between Eq.~\eref{eq:sigma_config} and Eq.~\eref{eq:sigma_charge} will be provided in Sec.~\ref{sec:Ar}.
The time-dependent differential scattering cross section from free electrons is given by
\begin{equation}
\frac{d^2 \sigma_\text{free}}{ d \Omega_F d \omega_F } ( t ) = \sum_{ E } P_{E}(t) \frac{ d^2 \sigma_{\text{free},E} }{ d \Omega_F d \omega_F },
\end{equation}
where $P_{E}(t)$ is given by Eq.~\eref{eq:P_E}.
$\sigma_{\text{free},E}$ will be derived in Sec.~\ref{sec:free_electron} [Eq.~\eref{eq:sigma_WH}].

Consequently, Eq.~\eref{eq:scattering_signal} may be evaluated with time-averaged configurational populations,
\begin{equation}\label{eq:S_config}
I_C(\Omega_F,\omega_F) = \mathcal{F} \left[ \sum_I \bar{P}_I \frac{ d^2 \sigma_{\text{bound},I} }{ d \Omega_F d \omega_F } + \sum_{E} \bar{P}_{E} \frac{ d^2 \sigma_{\text{free},E} }{ d \Omega_F d \omega_F } \right],
\end{equation}
or approximated with time-averaged charge-state populations,
\begin{equation}\label{eq:S_charge}
I_C(\Omega_F,\omega_F) \approx \mathcal{F} \left[ \sum_q \bar{P}_q \frac{ d^2 \sigma_{\text{bound},q} }{ d \Omega_F d \omega_F } + \sum_{E} \bar{P}_{E} \frac{ d^2 \sigma_{\text{free},E} }{ d \Omega_F d \omega_F } \right],
\end{equation}
where $\bar{P}_X$ is a pulse-weighted time-averaged quantity given by $\bar{P}_X = \int_{-\infty}^{\infty} dt \; g(t) P_X(t)$.
Note that the expression inside the brackets represents the effective Compton scattering cross section including ionization dynamics.
Equations~\eref{eq:S_config} and \eref{eq:S_charge} allow us to separate out the dynamical property $\bar{P}_X$, which depends on all relevant x-ray beam parameters, and the Compton scattering cross section $\sigma_X$, which depends on the atomic system and incident photon energy only.

\subsection{Doubly differential inelastic scattering cross section for bound electrons}\label{sec:bound_electron}

We employ nonrelativistic quantum electrodynamics based on the minimal coupling Hamiltonian and the Coulomb gauge~\cite{Santra09} to describe the light--matter interactions.
With this theoretical framework, nonresonant x-ray scattering is calculated using the $\bm{A}^2$ term in the interaction Hamiltonian, where $\bm{A}$ is the vector potential of the radiation field.

Let an incident x-ray photon be inelastically scattered by the bound electrons of the atomic system.
Let $(\omega_\text{in},\mathbf{k}_\text{in})$ be the incident photon energy and momentum, and $(\omega_F,\mathbf{k}_F)$ be the scattered photon energy and momentum.
Then, the inelastic part of the doubly differential scattering cross section (DDSCS) is given by (see detailed derivations in Ref.~\cite{Slowik14}),
\begin{align}\label{eq:DDSCS_full_angle}
\frac{ d^2 \sigma_\text{bound} }{ d \Omega_F d \omega_F } = 
\left( \frac{ d \sigma }{ d \Omega_F } \right)_\text{T} 
\frac{ \omega_F }{ \omega_\text{in} } 
\left[ 
\sum_f^\text{unocc.} \sum_i^\text{occ.} \delta( \varepsilon_f - \varepsilon_i + \omega_F -\omega_\text{in} ) \left| \int \! d^3 x \, \varphi_f^\dagger(\mathbf{x}) \varphi_i(\mathbf{x}) \rme^{\rmi \mathbf{Q} \cdot \mathbf{x} }\right|^2 
\right],
\end{align}
where the momentum transfer $\mathbf{Q}$ is defined by $\mathbf{Q} \equiv \mathbf{k}_\text{in} - \mathbf{k}_F$.
Here, $\varphi_p(\mathbf{x})$ is a spin-orbital, $\varepsilon_p$ is the associated orbital energy; $i$ runs over all orbitals occupied in the initial state of the atomic system, and $f$ runs over all unoccupied orbitals.
The Thomson scattering cross section, $\sigma_\text{T}(\Omega_F)$, for linearly polarized x-rays is given by
\begin{equation}\label{eq:TSCS}
\sigma_\text{T}(\Omega_F) = \left( \frac{d \sigma}{d \Omega_F} \right)_\text{T} = \alpha^4 ( 1 - \cos^2 \phi_F \sin^2 \theta_F ), 
\end{equation}
where $\theta_F$ is the polar angle and $\phi_F$ is the azimuthal angle of the scattered photon momentum with respect to the incident photon momentum and polarization axis. 
Here $\alpha$ is the fine-structure constant, and $\alpha^4$ corresponds to the square of the classical electron radius, $r_0^2 \approx 0.080$~barns.
Note that the expression inside the brackets in Eq.~\eref{eq:DDSCS_full_angle} does not depend on $\phi_F$, assuming that the electronic density of the atomic system is spherically symmetric.
Dependence on the azimuthal angle comes solely from the Thomson scattering cross section.
Therefore, we plot Compton spectra as a function of $\theta_F$ and $\omega_F$ after dividing Eq.~\eref{eq:DDSCS_full_angle} by Eq.~\eref{eq:TSCS},
\begin{equation}\label{eq:DDSCS}
S_\text{bound}( \theta_F, \omega_F ) = \left. \left( \frac{ d^2 \sigma_\text{bound} }{ d \Omega_F d \omega_F } \right) \middle/ \left( \frac{d \sigma}{d \Omega_F} \right)_\text{T} \right. .
\end{equation}

\begin{figure}
\includegraphics[]{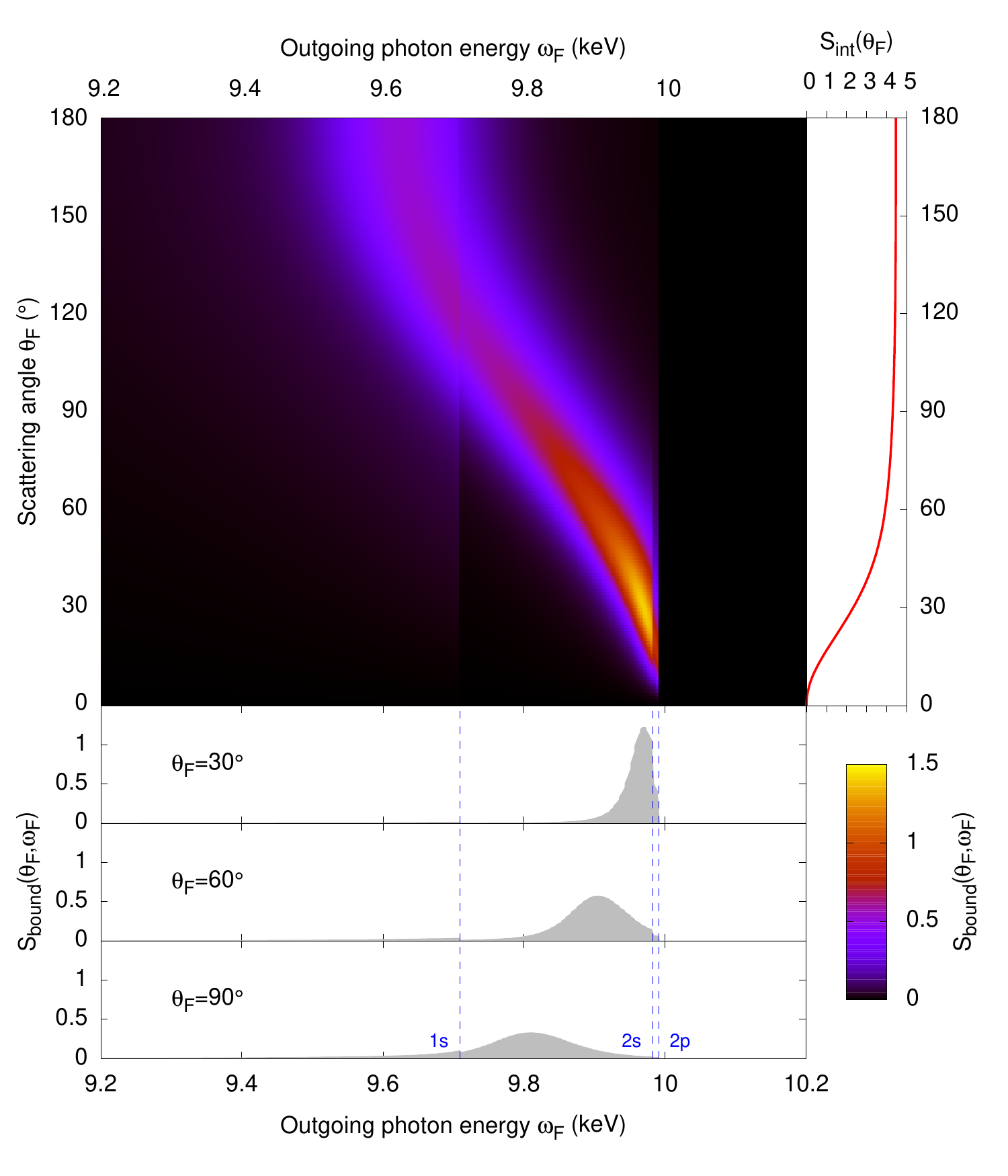}
\caption{\label{fig:C_10keV}Compton spectra for the ground-state configuration of neutral C interacting with 10-keV x-rays.}
\end{figure}

\Fref{fig:C_10keV} shows Compton spectra for the ground-state configuration of neutral carbon interacting with 10-keV x-rays.
The main panel shows a color map of $\sigma_\text{bound}( \theta_F, \omega_F )$ as a function of angle (y-axis) and energy (x-axis).
The three panels below the main panel show the DDSCS for specific angles ($\theta_F$=30\degree, 60\degree, and 90\degree).
One can see in \Fref{fig:C_10keV} the sum of the Compton profiles of individual spatial orbitals, reflecting their electron momentum distribution~\cite{Cooper85}.
In fact, the Compton profile~\cite{Eisenberger70} is defined by the expression inside the brackets in Eq.~\eref{eq:DDSCS_full_angle}.
The more contracted the spatial orbital, the more energetically extended is the Compton profile.
The vertical line around 9.7~keV and the two vertical lines near 10~keV correspond to $1s$-edge (291~eV), $2s$-edge (18~eV), and $2p$-edge (9~eV), respectively.
The edges are marked with blue dashed lines in the lower panels.
On the right panel, the angle- and energy-resolved Compton spectrum is integrated over the scattered photon energy in order to examine the angle-resolved Compton spectrum,
\begin{equation}\label{eq:angle-resolved}
S_\text{int}( \theta_F ) = \int \! d\omega_F \, S_\text{bound}( \theta_F, \omega_F ).
\end{equation}
Note that Eq.~\eref{eq:angle-resolved} provides the Compton scattering background without considering the energy resolution~\cite{Slowik14}.

\subsection{Doubly differential inelastic scattering cross section for free electrons}\label{sec:free_electron}

\begin{figure}
\includegraphics[]{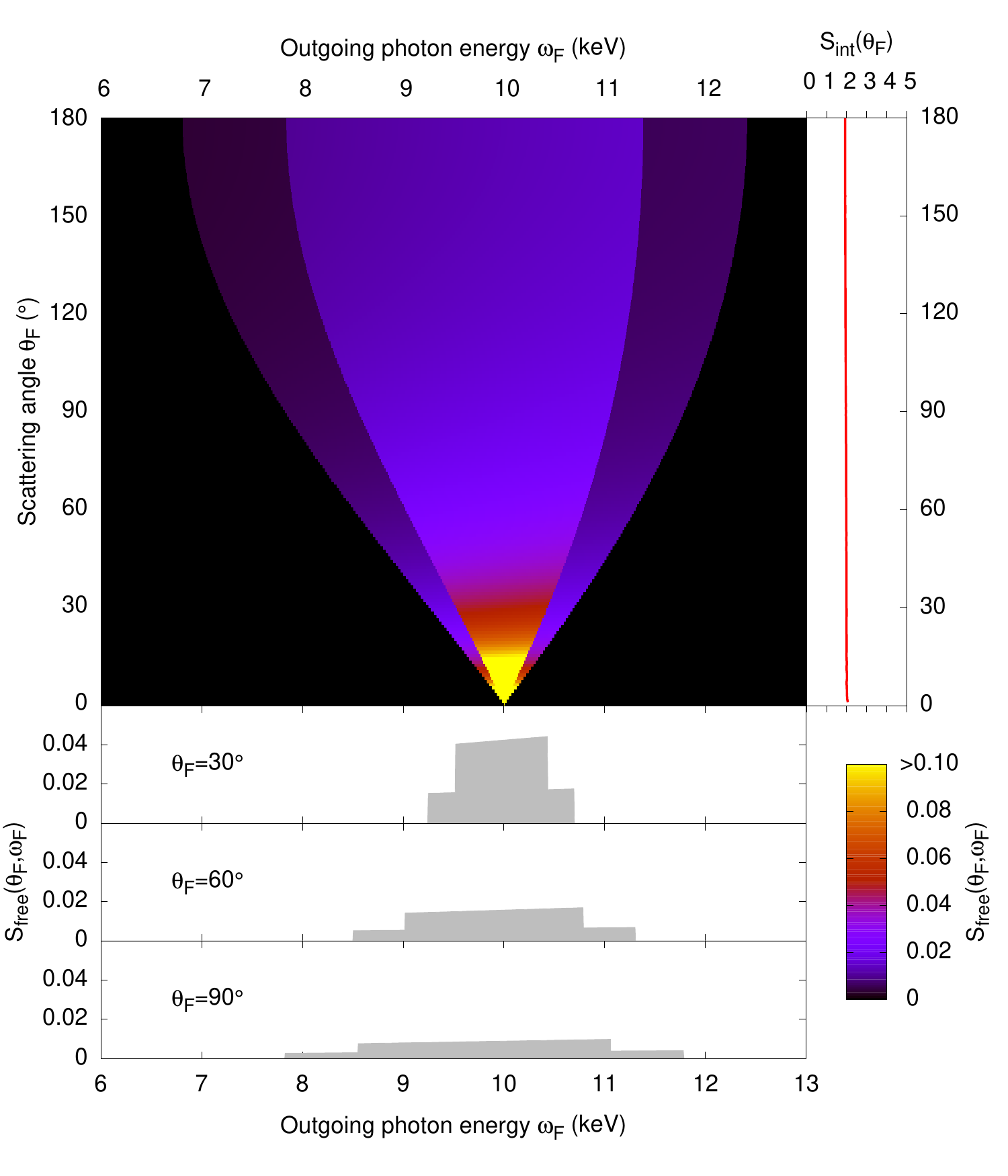}
\caption{\label{fig:elec_2keV_5keV}Compton spectra for two free electrons, one 2-keV electron and one 5-keV electron, interacting with 10-keV x-rays.}
\end{figure}

Now let an incident x-ray photon be inelastically scattered by a free electron that has a finite energy after ionization.
Let $\mathbf{p}_\text{in}$ be the initial momentum of the free electron and $\mathbf{p}_F$ be the final momentum.
From momentum conservation, $\mathbf{k}_\text{in} + \mathbf{p}_\text{in} = \mathbf{k}_F + \mathbf{p}_F$.
From energy conservation, $\omega_\text{in} + \left| \mathbf{p}_\text{in} \right|^2 / 2 = \omega_F + \left| \mathbf{p}_F \right|^2 / 2 $.
Thus, the outgoing photon energy is given by
\begin{equation}\label{eq:w_F}
\omega_F = \omega_\text{in} + \frac{ \left| \mathbf{p}_\text{in} \right|^2 }{ 2 } - \frac{\left| \mathbf{k}_\text{in} - \mathbf{k}_F + \mathbf{p}_\text{in} \right|^2}{2} 
= \omega_\text{in} - \frac{ Q^2 }{2} - \mathbf{Q} \cdot \mathbf{p}_\text{in}.
\end{equation}
We apply the Waller-Hartree approximation ($\omega_F \approx \omega_\text{in}$ on the right-hand side of Eq.~\eref{eq:w_F}) to obtain a simple expression.
The deviation due to the Waller-Hartree approximation will be discussed later.
Thus, the Compton shift, $Q^2 / 2$, is expressed by
\begin{equation}\label{eq:Compton_shift}
\frac{Q^2}{2} = \frac{\alpha^2}{2} \left( \omega_F^2 + \omega_\text{in}^2 - 2 \omega_\text{in} \omega_F \cos \theta_F \right)
\approx \alpha^2 \omega_\text{in}^2 \left( 1  - \cos \theta_F \right),
\end{equation} 
and the Doppler shift, $\mathbf{Q} \cdot \mathbf{p}_\text{in}$, is expressed by
\begin{equation}\label{eq:Doppler_shift}
\mathbf{Q} \cdot \mathbf{p}_\text{in} = Q \left| \mathbf{p}_\text{in} \right| \cos \theta' 
\approx 2 \alpha \omega_\text{in} \sin \! \left( \theta_F / 2 \right) \sqrt{ 2 \varepsilon_{\mathbf{p}_\text{in}} } \cos \theta',
\end{equation}
where $\varepsilon_{\mathbf{p}_\text{in}}$ is the initial kinetic energy of the free electron ($ = \left| \mathbf{p}_\text{in} \right|^2 / 2$) and $\theta'$ is the angle between $\mathbf{p}_\text{in}$ and $\mathbf{Q}$.
Then the outgoing photon energy is simply given by
\begin{equation}\label{eq:omega_F}
\omega_F = \omega_\text{in} - \alpha^2 \omega_\text{in}^2 ( 1 - \cos \theta_F ) - 2 \alpha \omega_\text{in} \sqrt{ 2 \varepsilon_{\mathbf{p}_\text{in}} } \sin \! \left( \theta_F / 2 \right) \cos \theta'.
\end{equation}

We assume that the free electron is described by a plane wave state. 
With this assumption, the summation over $f$ in Eq.~\eref{eq:DDSCS_full_angle} is converted into an integral over $\mathbf{p}_F$, and the summation over $i$ into an integral over $\mathbf{p}_\text{in}$~\cite{Slowik_phdthesis}.
The DDSCS then goes over into
\begin{align}\label{eq:sigma_free}
\frac{ d^2 \sigma_\text{free} }{ d \Omega_F d \omega_F } 
&= 
\left( \frac{ d \sigma }{ d \Omega_F } \right)_\text{T} 
\frac{ \omega_F }{ \omega_\text{in} } 
\int \! d^3 p_F \int \! d^3 p_\text{in} \;
f(\mathbf{p}_\text{in})
\delta( \varepsilon_{\mathbf{p}_F} - \varepsilon_{\mathbf{p}_\text{in}} + \omega_F -\omega_\text{in} ) 
\delta( \mathbf{Q} + \mathbf{p}_\text{in} - \mathbf{p}_F )
\nonumber
\\
&= 
\left( \frac{ d \sigma }{ d \Omega_F } \right)_\text{T} 
\frac{ \omega_F }{ \omega_\text{in} } 
\int \! d^3 p_\text{in} \;
f(\mathbf{p}_\text{in})
\delta( \omega_F - \tilde{\omega}_F(\mathbf{p}_\text{in}) ) ,
\end{align}
where $\tilde{\omega}_F$ is given in Eq.~\eref{eq:w_F} as a function of $\mathbf{p}_\text{in}$, and $f(\mathbf{p}_\text{in})$ is the initial momentum distribution.
Let us assume that the angular distribution of the ionized electron with a given kinetic energy $E$ is isotropic, 
\begin{equation}
f(\mathbf{p}_\text{in}) = \frac{1}{8 \pi E} \delta \left( | \mathbf{p}_\text{in} | - \sqrt{ 2 E } \right),
\end{equation}
where the kinetic energy $E$ is determined by the photoionization or Auger (Coster-Kronig) decay process that has produced the free electron.
With this isotropic $f(\mathbf{p}_\text{in})$, the integral over $\mathbf{p}_\text{in}$ in Eq.~\eref{eq:sigma_free} is readily evaluated as
\begin{equation}\label{eq:sigma_WH}
\frac{ d^2 \sigma_{\text{free},E} }{ d \Omega_F d \omega_F } = 
\left( \frac{ d \sigma }{ d \Omega_F } \right)_\text{T} \frac{ \omega_F }{ \omega_\text{in} } \cdot \frac{1}{ 4 \alpha \omega_\text{in} \sin \! \left( \theta_F / 2 \right) \sqrt{ 2 E } },
\end{equation}
for a range of $\omega_F$ that is given by Eq.~\eref{eq:omega_F} for $0 \leq \theta' \leq \pi$,
\begin{equation}\label{eq:omega_F_range}
\left| \omega_F - \left\lbrace \omega_\text{in} - \alpha^2 \omega_\text{in}^2 ( 1 - \cos \theta_F ) \right\rbrace \right| \leq 2 \alpha \omega_\text{in} \sqrt{ 2 E } \sin \! \left( \theta_F / 2 \right).
\end{equation}
After dividing by the Thomson scattering cross section, we obtain an angle-resolved and energy-resolved Compton spectrum for a free electron with kinetic energy $E$,
\begin{equation}\label{eq:S_WH}
S_{\text{free},E}(\theta_F,\omega_F)
= \left. \left( \frac{ d^2 \sigma_{\text{free},E} }{ d \Omega_F d \omega_F } \right) \middle/ \left( \frac{d \sigma}{d \Omega_F} \right)_\text{T} \right..
\end{equation}

In Eq.~\eref{eq:omega_F_range}, both the Compton shift, $\alpha^2 \omega_\text{in}^2 ( 1 - \cos \theta_F )$, and the Doppler shift, $2 \alpha \omega_\text{in} \sqrt{ 2 E } \sin \! \left( \theta_F / 2 \right)$, are obtained within the Waller-Hartree approximation [cf.\ Eqs.~\eref{eq:Compton_shift} and \eref{eq:Doppler_shift}].
The Compton shift is proportional to $\omega_\text{in}^2$ and the Doppler shift is proportional to $\omega_\text{in}$ and $\sqrt{ E }$.
Also note that the Compton shift is always a redshift (a down shift in energy), whereas the Doppler shift shows both blueshift and redshift (up and down shifts in energy).
Therefore, the Waller-Hartree approximation employed here has to be reconsidered when high-energy x-ray photons and high-energy free electrons are involved.
For example, x-rays with a photon energy of 10~keV could make a Compton shift of $\sim$400~eV.
If x-rays with the same incident photon energy are scattered by free electrons with a kinetic energy of 5~keV, then they could make a Doppler shift of 2.8~keV.
 
To test the validity of the Waller-Hartree approximation, we numerically solve $\omega_F$ in Eq.~\eref{eq:w_F} without the approximation as follows.
Equation~\eref{eq:w_F} represents a nonlinear equation for $\omega_F$ because $Q$ on the right-hand side contains $\omega_F$ via Eqs.~\eref{eq:Compton_shift} and \eref{eq:Doppler_shift} without the approximation.
Let us consider the following coupled equations from Eqs.~\eref{eq:w_F}, \eref{eq:Compton_shift}, and \eref{eq:Doppler_shift} in an iterative way:
\begin{align}
Q^{(k)} &= \alpha \sqrt{ {\omega_{F}^{(k-1)}}^2 + \omega_\text{in}^2 - 2 \omega_\text{in} {\omega_{F}^{(k-1)}} \cos \theta_F },
\\
{\omega_{F}^{(k)}} &= \omega_\text{in} - \frac{{Q^{(k)}}^2}{2} - Q^{(k)} \sqrt{ 2 E } \cos \theta'.
\end{align}
At the $k$th step, $Q^{(k)}$ is calculated using $\omega_{F}^{(k-1)}$ from the previous step, and then $\omega_{F}^{(k)}$ is calculated using $Q^{(k)}$, and so on.
This iterative procedure starts with Eq.~\eref{eq:omega_F} for $\omega_F^{(0)}$, which is the solution employing the Waller-Hartree approximation, and repeats until $\omega_F$ is converged.
For the above example ($\omega_\text{in}$=10~keV and $E$=5~keV), the converged results in this case show deviations less than 8\% from those obtained from the Waller-Hartree approximation.

\Fref{fig:elec_2keV_5keV} shows the Compton spectra for two free electrons with kinetic energies of 2~keV and 5~keV, respectively, interacting with 10-keV x-rays.
All calculations are done with the Waller-Hartree approximation.
Due to the Doppler shift by free electrons with arbitrary directions, the shift in $\omega_F$ can be negative (redshift) for $\theta' < \pi/2$ or positive (blueshift) for $\theta' > \pi/2$.
Because of the high electron kinetic energies assumed, the shifts are quite large and the Compton spectra are spread out over a broad range.
For the 5-keV case, the shifts become $\sqrt{5/2} \sim 1.6$ times larger than for the 2-keV case, but the height is reduced accordingly.
Note that the energy-integrated Compton spectrum on the right panel gives 2$\sigma_T(\Omega_F)$ from two free electrons, showing no dependence on the scattering angle.
It is also numerically confirmed that the energy-integrated cross section for each free electron gives 1 in units of the Thomson scattering cross section, $\sigma_T(\Omega_F)$.

\subsection{Efficient evaluation of spherical Bessel functions}\label{sec:Bessel}

\begin{figure}
\includegraphics[]{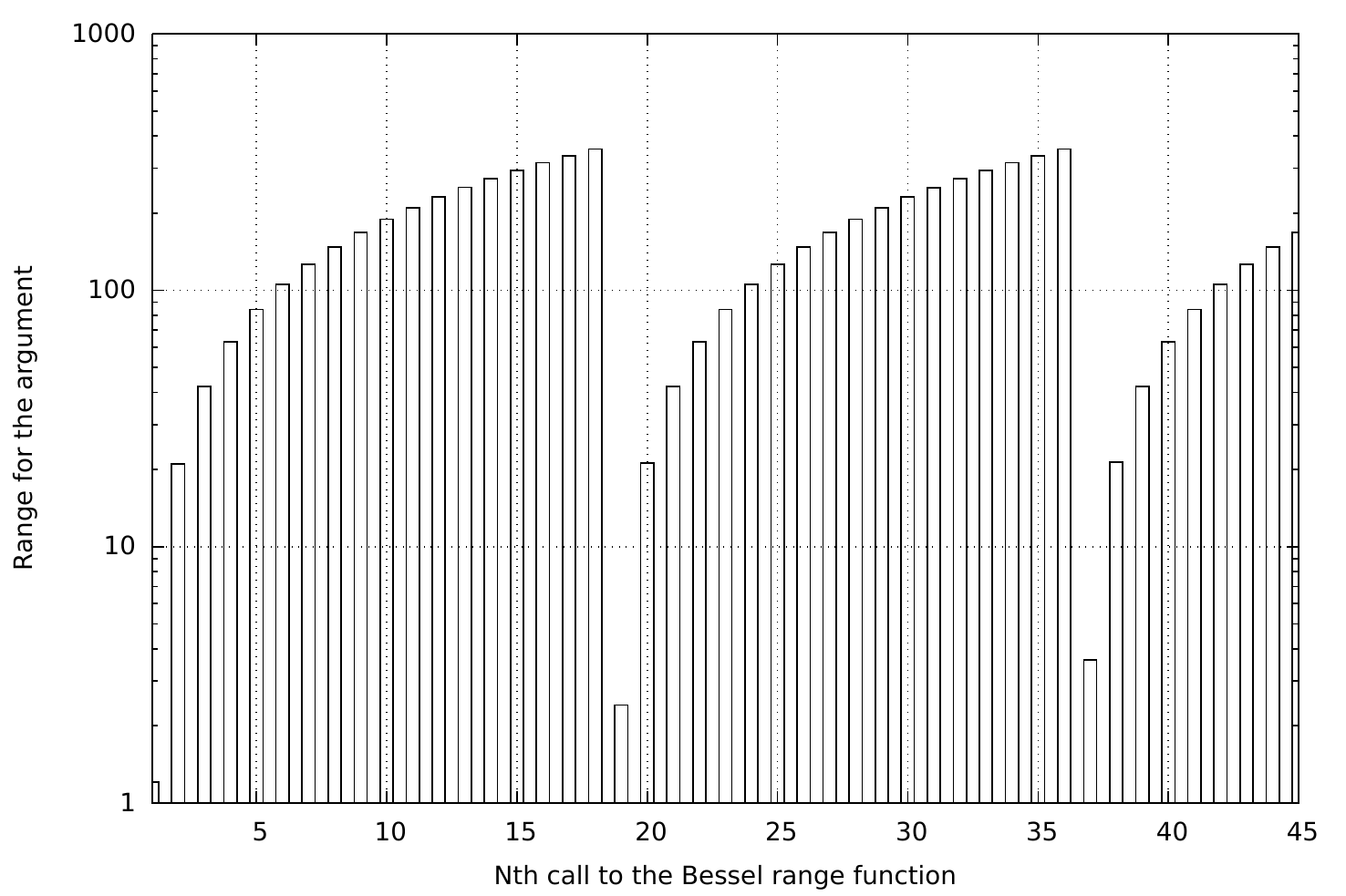}
\caption{\label{fig:intervals}Argument range in a sequence of the Bessel range function calls.}
\end{figure}
\begin{figure}
\includegraphics[]{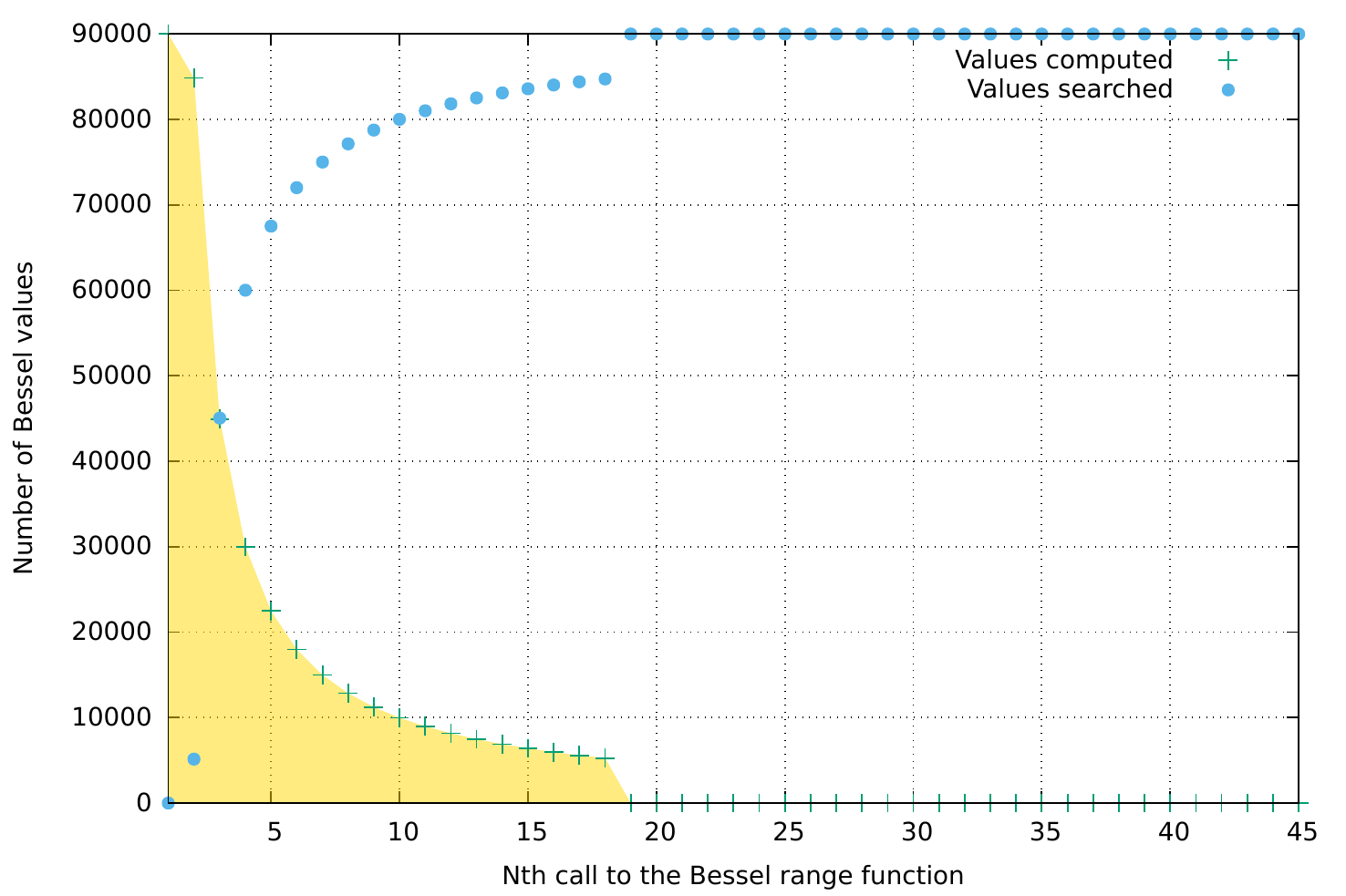}
\caption{\label{fig:computation}Number of computed and searched values in a sequence of the Bessel range function calls.}
\end{figure}
\begin{figure}
\includegraphics[]{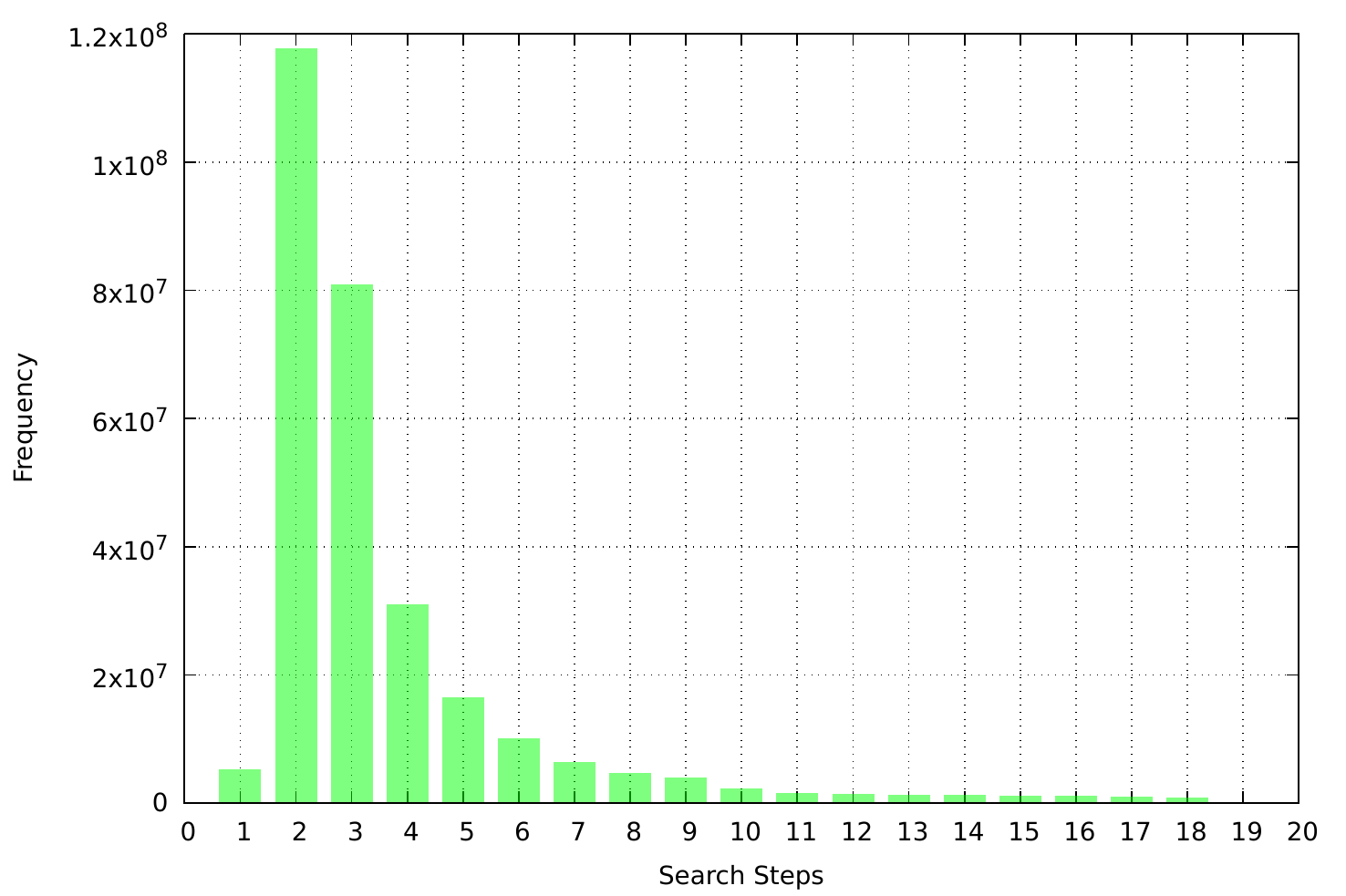}
\caption{\label{fig:search_cycles}Histogram of search steps used in the Bessel range function calculations.}
\end{figure}

The overall runtime for the DDSCS calculations of a heavy atom like Xe is dominated by evaluations of the Bessel functions.
We implement an efficient scheme to evaluate the spherical Bessel function of the $j$-type~\cite{NIST10} in a Fortran-90 function.
The code is based on the mathematical work by Lentz~\cite{Lentz90}. 
In the following we just call it the Bessel function. 
Since we need the evaluation of the Bessel function for different $l$ from 0 to $l_\text{max}$ and $x$ from 0 to $x_\text{max}$, where $x=Qr$ on a radial grid, we code this functionality in the so called Bessel range function.
The Bessel range function calls the Bessel function for actual computations, 
but at the same time it avoids to do so if an interpolation between two already computed Bessel values is possible.

Here we describe how our implementation works in detail by showing the following three figures.
These figures are based on a single DDSCS calculation of the neutral ground-state Xe atom at an incoming photon energy of 10~keV, with 18 different scattering angles, $l_\text{max}=35$, outgoing photon energies from 7 to 10~keV, an energy step of 10~eV, and 90,000 points on the uniform radial grid ($0 \leq r \leq 450$~a.u.) for the electron continuum states. 
The same parameters are used in the Compton spectra calculations of Ar and Xe in the Results (Sec.~\ref{sec:results}), except for 181 scattering angular points to get better angular resolution.

In \Fref{fig:intervals} we plot the argument range at the $N$-th call to the Bessel range function. 
For the first 18 calls the argument range ramps up, and it increases in a similar fashion for the next calls.
Such block with 18 calls corresponds to the 18 different scattering angles at fixed outgoing photon energy.
So it is possible to compute spherical Bessel functions only up to a maximum argument, to store arguments and results in a table, and then to search an argument in the table and perform a linear interpolation to obtain a resulting Bessel value. 
That means that actual computing is necessary only for a very small fraction of calls, which is illustrated in \Fref{fig:computation}.
\Fref{fig:computation} plots the number of computed and searched values in a sequence of the Bessel range function calls.
It shows that only the first ramp in \Fref{fig:intervals} includes computation of spherical Bessel functions.
In the first ramp, we apply the optimization scheme as follows.
At the first call to the Bessel range function, one computes every single function value.
In all the following calls, one looks up the argument and interpolates until it does not fall into the previous range anymore, and then one has to actually compute a value. 

As shown in \Fref{fig:computation}, most of the time may be used for searching, rather than actual computation of the Bessel function.
Thus, the searching algorithm is another key to optimize this calculation procedure.
A general purpose method such as the binary search algorithm would give a computation complexity of $O(\log_2 n)$, where $n$ is the number of data.
In our case we have accumulated a table of 394,037 values, which would lead to about 19 search steps on average by using the binary search algorithm.
\Fref{fig:search_cycles} shows a histogram of search steps in our simulation.
Our searching algorithm exploits that the Bessel function arguments to be searched are in ascending order.
Therefore, it requires very often only two search steps, and about 4 steps on average.

With this overall optimization scheme we can achieve about 8 times faster calculations than those previously used in Compton scattering calculations~\cite{Slowik14}.
The runtime on a single CPU (Intel Xeon E5-2609) is reduced from 27h 6min to 3h 18min, for a DDSCS calculation of the neutral ground-state Xe atom with 181 scattering angular points.
The results before and after optimization are identical to within 0.1\% difference.
Also it is worthwhile to note that the new Bessel function routine can handle a very high $l$ $( > 100)$, so it can accommodate a very high energy regime.
The more extreme the parameters get, the bigger are the advantages of our optimization.

\section{Results}\label{sec:results}

\subsection{Compton spectra of atomic Ar at an incident photon energy of 10~keV}\label{sec:Ar}
\begin{figure}
\includegraphics[]{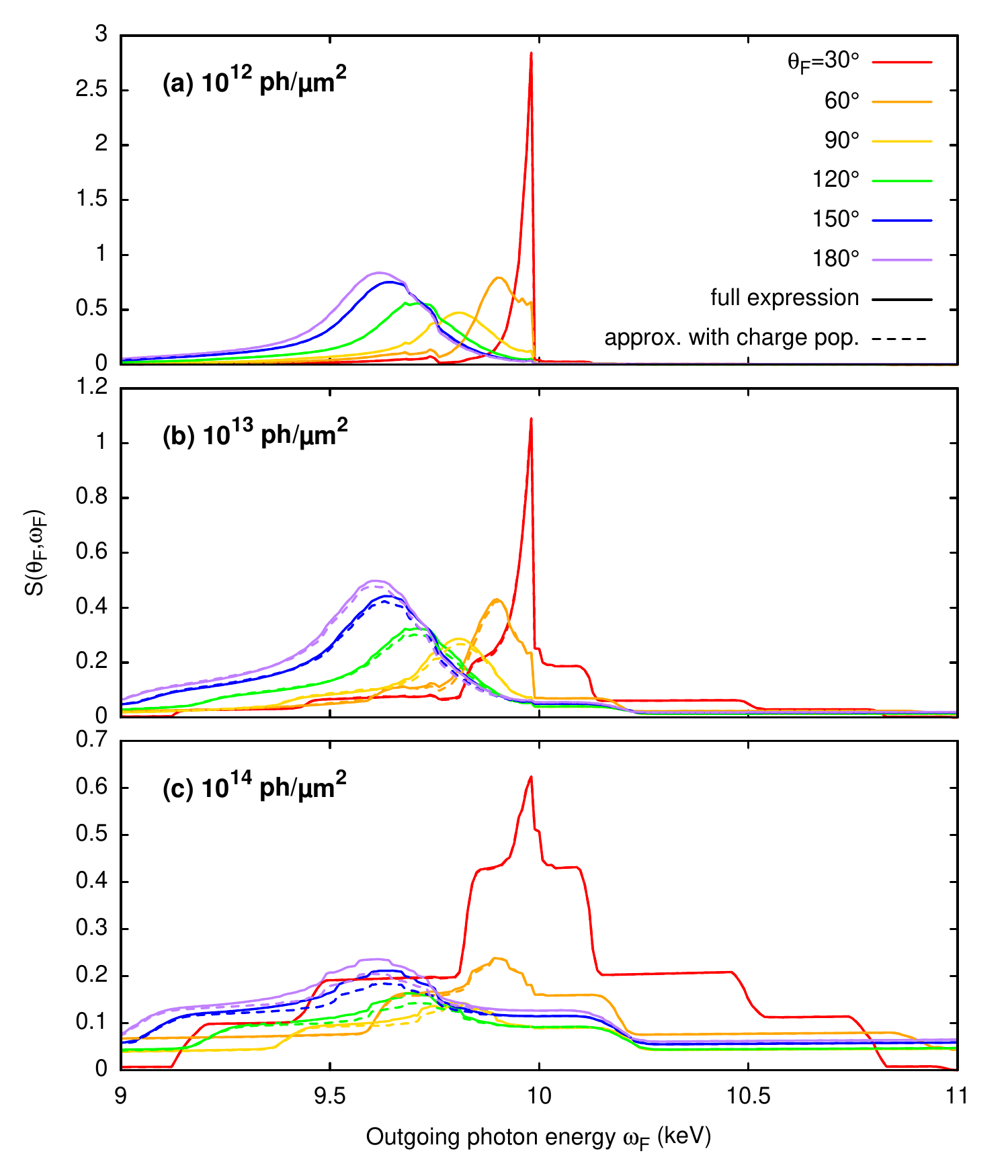}
\caption{\label{fig:Ar_compare}Comparison of the Compton spectra of Ar calculated using configuration populations [full expression of Eq.~\eref{eq:S_config}, plotted with solid curves] and using charge-state populations [approximated expression of Eq.~\eref{eq:S_charge}, plotted with dashed curves].
Note that in the weak-field regime, shown in panel~(a), the full expression and the approximated expression give virtually identical results.
Different colors indicate different scattering angles.}
\end{figure}

First, we compare the Compton scattering cross section from bound electrons, calculated using the full expression based on configuration populations [Eq.~\eref{eq:S_config}] and using the approximation based on charge-state populations [Eq.~\eref{eq:S_charge}].
We plot the angle- and energy-resolved Compton spectra after dividing Eq.~\eref{eq:S_config} or Eq.~\eref{eq:S_charge} by the incident x-ray fluence and the Thomson scattering cross section [Eq.~\eref{eq:TSCS}],
\begin{equation}
S(\theta_F,\omega_F) = \left. \left[ \frac{I_C(\Omega_F,\omega_F)}{\mathcal{F}} \right] \middle/ \sigma_\text{T}(\Omega_F) \right..
\end{equation}
\Fref{fig:Ar_compare} shows the Compton spectra of argon at different scattering angles shown with different colors and several fluences shown in different panels: (a) $10^{12}$~ph/$\mu$m$^2$, (b) $10^{13}$~ph/$\mu$m$^2$, and (c) $10^{14}$~ph/$\mu$m$^2$.
The one-photon absorption saturation fluence of Ar at 10~keV is $2.4\times10^{12}$~ph/$\mu$m$^2$.
The temporal pulse shape is a Gaussian with a 30-fs full width at half maximum (FWHM).
The solid curves indicate the full expression, whereas the dashed curves reflect the approximated expression.
One can see no visible differences between the solid and dashed curves at a fluence of $10^{12}$~ph/$\mu$m$^2$, which is less than the one-photon absorption saturation fluence, i.e., in the weak intensity regime.
Also at low angles (for example, $\theta_F$=30\degree\ and 60\degree), there is almost no deviation between the two approaches.
The deviation increases when the fluence is increased and the scattering angle is higher.
Since the two different approaches matter for bound electrons only, the discrepancy between them will be somewhat reduced as the fluence is increased further, where the contribution from ionized electrons becomes dominant.

\begin{figure}
\includegraphics[]{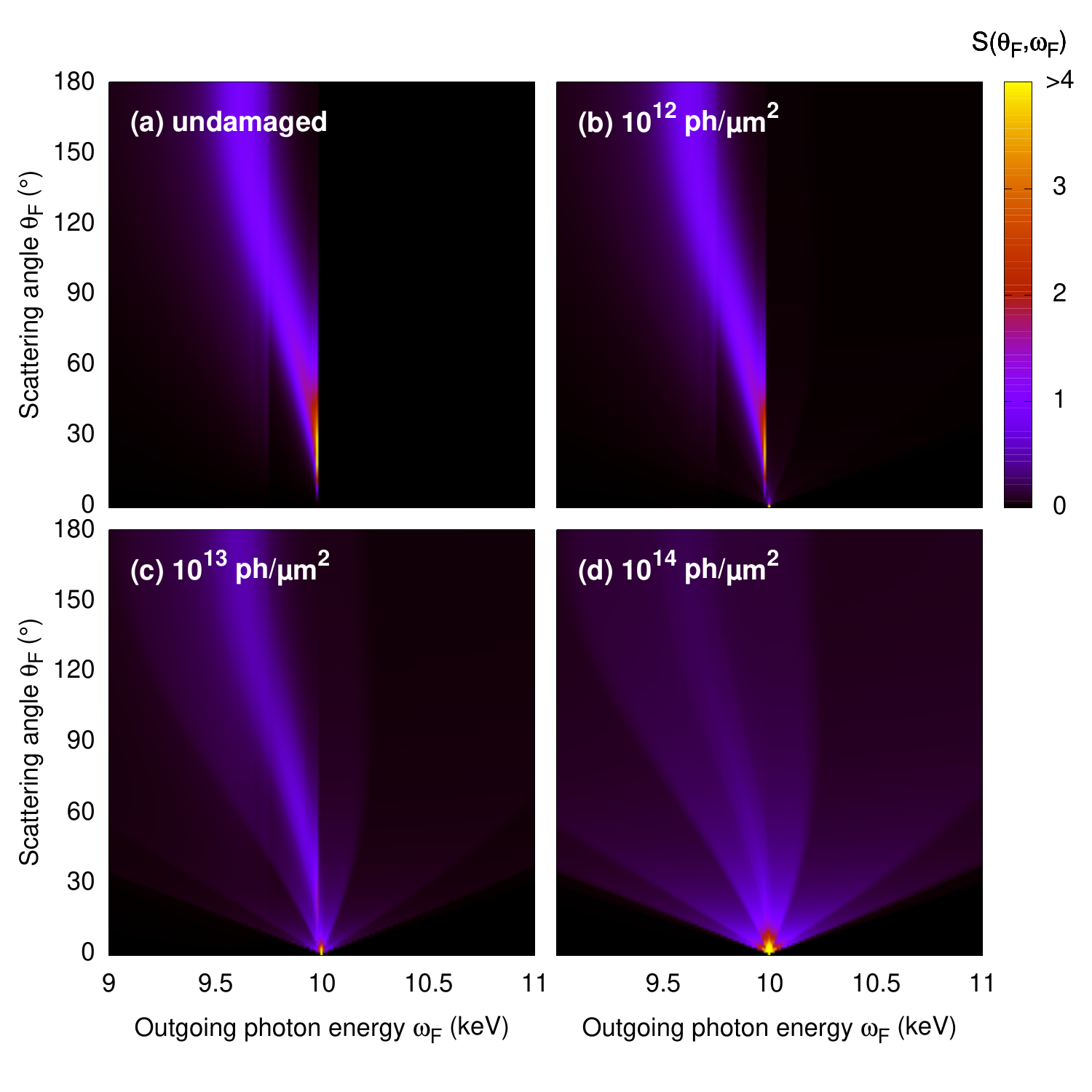}
\caption{\label{fig:Ar_fluences}Compton spectra of Ar at 10 keV for several fluences.}
\end{figure}

In \Fref{fig:Ar_fluences}, we plot the Compton spectra of argon in a color map as a function of angle and energy.
Panel (a) represents the neutral argon case (undamaged), assuming that no ionization dynamics occur when interacting with an extremely weak x-ray pulse.
The other panels reflect ionization dynamics induced by intense x-ray pulses.
The x-ray fluence used is $10^{12}$~ph/$\mu$m$^2$ in panel (b), $10^{13}$~ph/$\mu$m$^2$ in panel (c), and $10^{14}$~ph/$\mu$m$^2$ in panel (d).
The average charge state at the end of the ionization dynamics is (b) $+1.5$, (c) +10, and (d) +17, respectively.
Panel (b) shows the weak intensity case (less than the one-photon absorption saturation fluence), and there are almost no differences from the neutral argon case, except for faint blueshifted signals at low angles due to a few contributions from ionized electrons.
In panels (a) and (b), one can see characteristic features of the momentum distribution of the bound electrons in Ar: the vertical lines near 9.7~keV and 10~keV corresponding to $L$-shell and $M$-shell edges, respectively.
Panels (c) and (d) demonstrate the high intensity case.
The bound-electron features diminish and the free-electron contributions emerge.
Different contributions from various electron kinetic-energies stack up on top of the bound-electron contribution [for example, the stair-like structure of the red curve in \Fref{fig:Ar_compare}(c)].
The Doppler shift from free electrons is proportional to $\sin(\theta_F / 2)$.
Therefore signals at high angles are spread out in a wide energy range, while signals at low angles become concentrated on a blob.
Both panels (c) and (d) clearly demonstrate that Compton spectra are deformed in the high-intensity regime, due to large contributions from free electrons.

\subsection{Compton spectra of atomic Xe at an incident photon energy of 10~keV}\label{sec:Xe}
\begin{figure}
\includegraphics[]{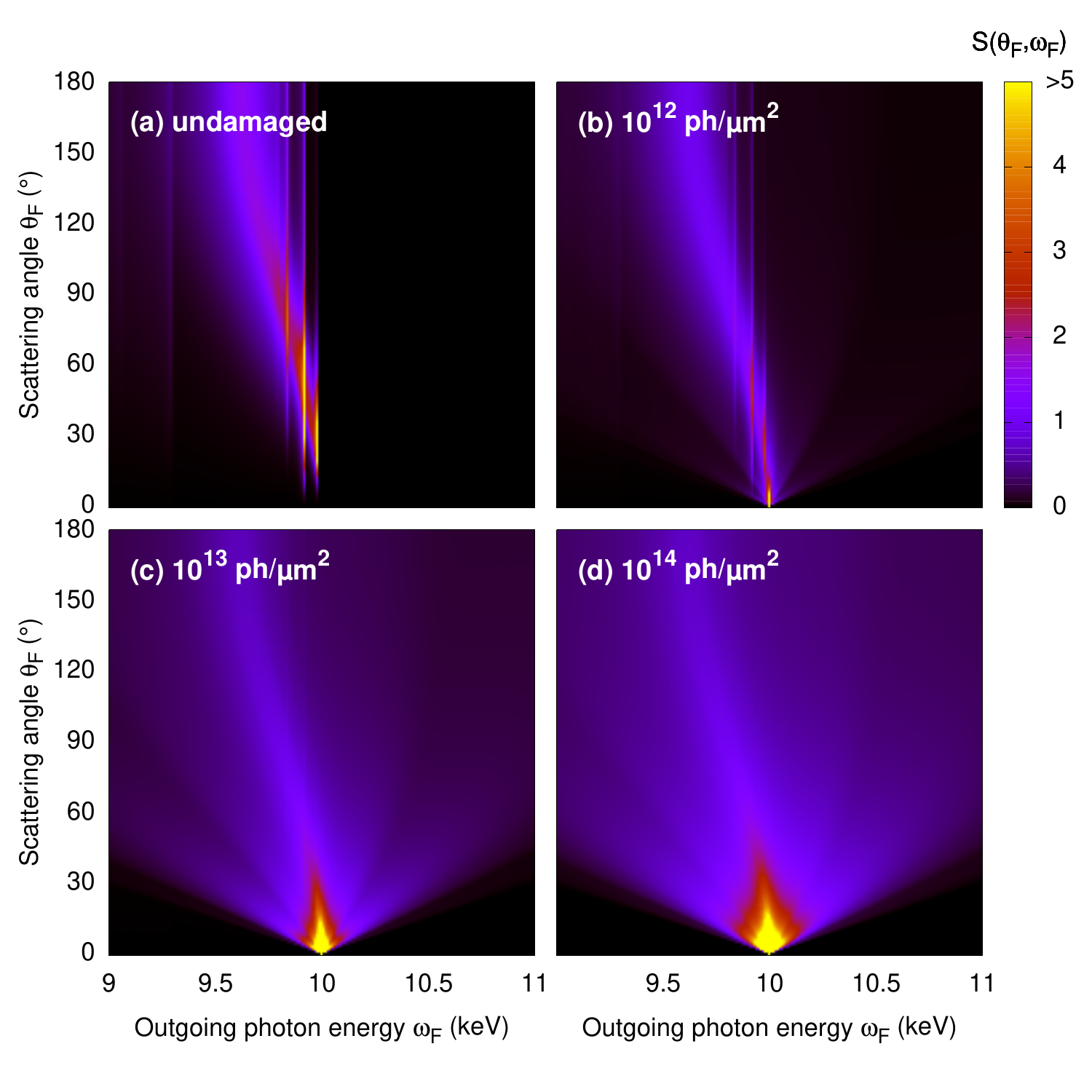}
\caption{\label{fig:Xe_fluences}Compton spectra of Xe at 10~keV for several fluences.}
\end{figure}

Equation~\eref{eq:S_charge} enables us to calculate the Compton scattering differential cross section for heavy atoms like Xe.
Heavy atoms undergo more severe ionization than light atoms by interacting with intense x-ray pulses, because photoionization cross sections are higher and there are many electrons that may be ionized via deep inner-shell ionization cascades~\cite{Rudek12,Fukuzawa13}.
\Fref{fig:Xe_fluences} shows the Compton spectra of xenon in a color map as a function of angle and energy.
Panel (a) represents the neutral xenon case (undamaged) and other panels show the high intensity cases with the same fluences used in \Fref{fig:Ar_fluences}.
The pulse duration used is 30~fs FWHM.
Note that the one-photon absorption saturation fluence of Xe at 10~keV is $2.9\times10^{11}$~ph/$\mu$m$^2$.
Therefore x-ray multiphoton dynamics become dominant with all the fluences used in \Fref{fig:Xe_fluences}.
The average charge state at the end of the ionization dynamics is (b) $+15$, (c) +49, and (d) +52, respectively.
In the last case, all electrons except $1s$ are fully ionized after the pulse.
In panels (a) and (b), one can see sharp vertical lines at $M$-, $N$-, and $O$-shell edges.
The strong peak of the redshifted signal (from 10~keV toward the left) indicates the momentum distribution of the bound electrons.
On the other hand, this peak is smeared out when the fluence is increased, as shown in panels (c) and (d).
In contrast to the Ar case in \Fref{fig:Ar_fluences}, the free-electron contributions are seemingly added up smoothly.

\begin{figure}
\includegraphics[]{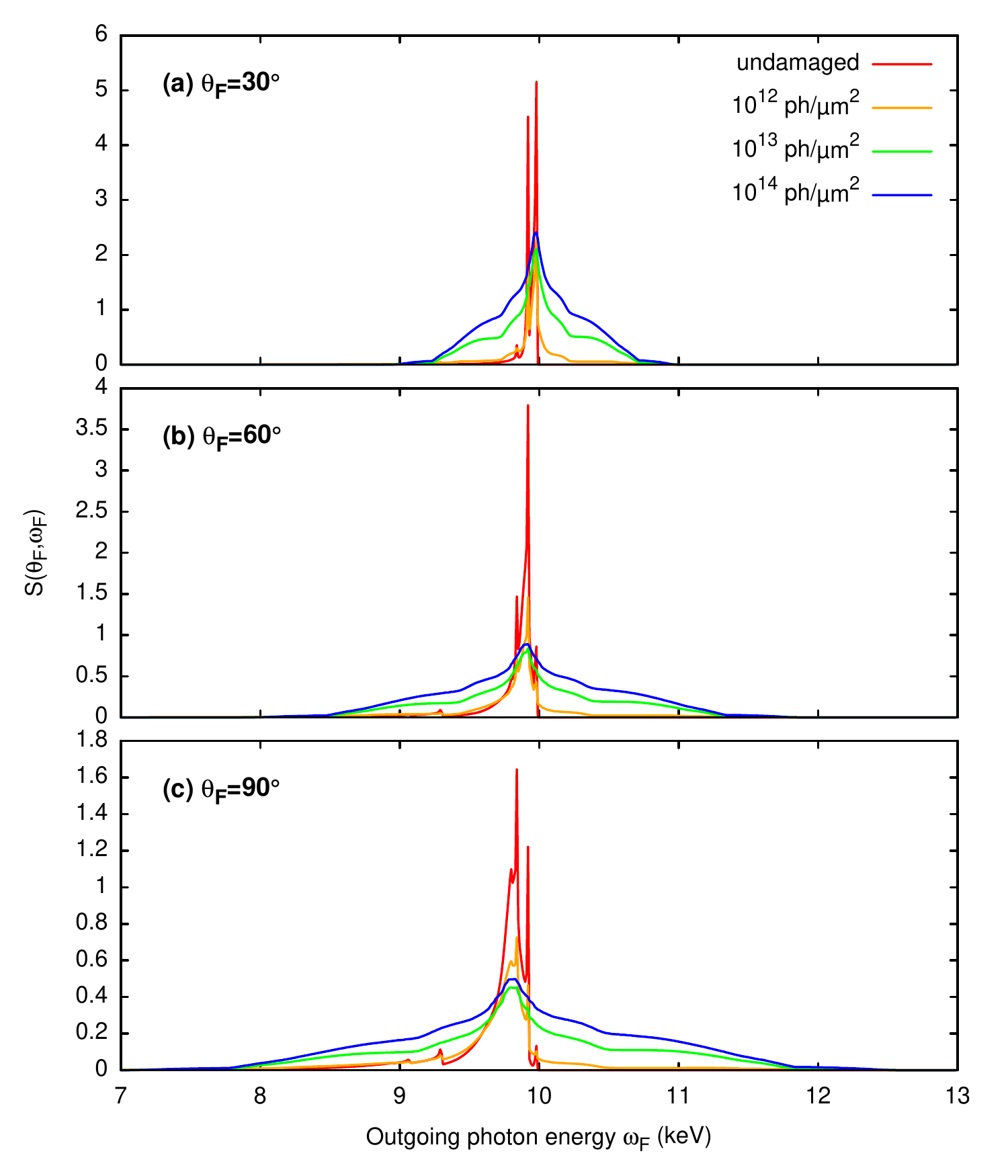}
\caption{\label{fig:Xe_angles}Compton spectra of Xe at 10~keV for several fluences and scattering angles.}
\end{figure}

To observe features more clearly, we plot the Compton spectra for several scattering angles and fluences in \Fref{fig:Xe_angles}.
Different panels correspond to different scattering angles: (a) 30\degree, (b) 60\degree, and (c) 90\degree.
Different colors indicate different fluences.
The red curve is the undamaged case, where one can see the characteristic structures from the bound-electron contribution.
The green and blue curves show that a very broad peak emerges because of many ionized electrons.
Even though the Compton spectra shown were calculated with charge-state populations [Eq.~\eref{eq:S_charge}], all configurations were taken into consideration in the ionization dynamics calculations.
Consequently, there is a large number of photoelectron and Auger-electron kinetic-energy spectral lines, giving rise to very broad and smooth contributions to the Compton scattering signal as shown in \Fref{fig:Xe_angles}.

\begin{figure}
\includegraphics[]{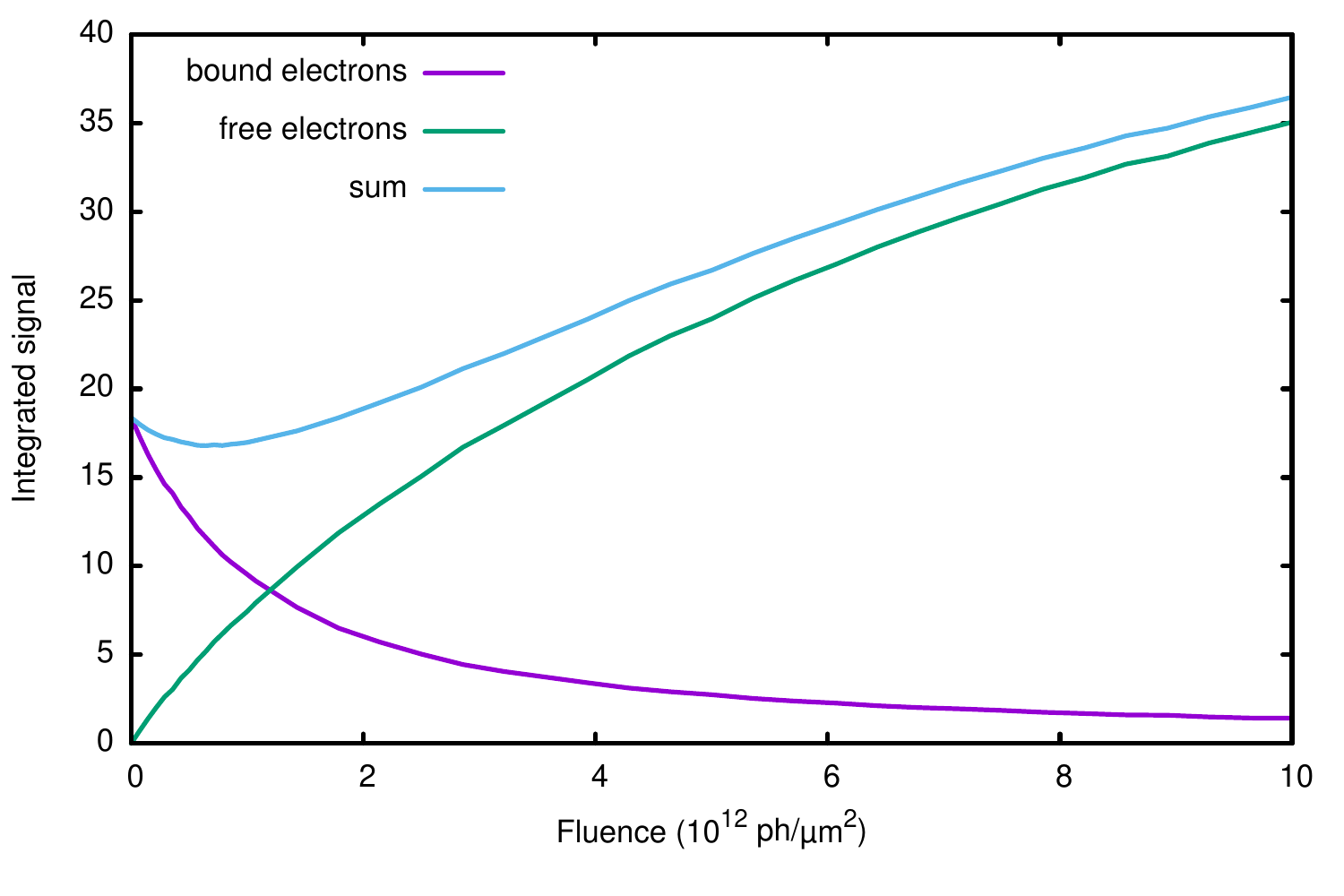}
\caption{\label{fig:Xe_all_integrated}Integrated Compton scattering signal of Xe at 10~keV as a function of fluence.}
\end{figure}

To further investigate the dependence of Compton spectra on the fluence, \Fref{fig:Xe_all_integrated} shows the total Compton scattering signal of Xe at 10~keV integrated over angles and energies, as a function of fluence.
The integrated signal is defined by
\begin{equation}
\tilde{S} = \left. \left[ \int \! d\Omega_F \int \! d\omega_F \, \frac{I_C(\Omega_F,\omega_F)}{\mathcal{F}} \right] \middle/ \tilde{\sigma}_T\right.,
\end{equation}
where the total Thomson scattering cross section is $\tilde{\sigma}_T = \int \! d\Omega_F \, \sigma_T(\Omega_F) = 8\pi r_0^2 / 3 \approx 0.67$~barns.
Note that $I_C/\mathcal{F}$ represents the effective cross section.
The purple line indicates the bound-electron contribution, while the green line represents the free-electron contribution.
The former decreases and the latter increases nonlinearly as the fluence increases, because of x-ray multiphoton ionization.
A single free electron contributes a cross section of the order of 1 in units of $\tilde{\sigma}_T$, as discussed in Sec.~\ref{sec:bound_electron}.
However, each electron bound to an atom may contribute less than 1.
For neutral Xe atom at 10~keV, the integrated signal from 54 bound electrons is $\sim 18.4 \tilde{\sigma}_T$ as shown at zero fluence in \Fref{fig:Xe_all_integrated}.
Therefore, the sum of bound- and free-electron contributions to the Compton scattering signal (shown with the blue line) keeps nonlinearly increasing as the fluence increases, even though the sum of bound- and free-electron numbers remains the same.

\section{Conclusion}\label{sec:conclusion}

In this paper, we have presented theoretical results for the Compton spectra of atoms at high x-ray intensity.
Interacting with intense x-ray pulses, an atomic system undergoes massive ionization, described by sequential multiphoton ionization dynamics.
Therefore a single atom produces a large number of free electrons and becomes a highly charged ion.
In contrast to the low x-ray intensity regime, where Compton scattering from bound electrons in neutral atoms is only considered, 
it is necessary in the high x-ray intensity regime to include the Compton scattering contributions from bound electrons in ions and from free electrons.
We have demonstrated that Compton spectra at high x-ray intensity are considerably deformed from those of neutral atoms, because of the contributions from ions and free electrons.
Due to individual ion contributions, characteristic peak structures of the Compton spectra are smoothed out.
As a consequence of large contributions from free electrons, the scattering signals exhibit not only a redshift but also a blueshift.
The effect becomes drastic as the x-ray fluence increases beyond the one-photon absorption saturation fluence, and it becomes more severe when a heavier atom is considered.
In order to make the Compton spectra calculations for heavy atoms feasible, we have introduced a simplified expression for bound electrons and proposed an efficient way to calculate spherical Bessel functions extensively used in the Compton calculations.
These present implementations have been incorporated into the extended \textsc{xatom} toolkit.
\textsc{xatom} augmented with Compton scattering enables us to investigate elastic and inelastic scattering dynamics on the same footing, including severe radiation damage when the sample is exposed to intense XFEL pulses.
Our present results can be utilized for interpreting single-particle molecular imaging experiment~\cite{Schlichting12} and for diagnosing warm dense matter generated by XFEL~\cite{Glenzer16}.

\ack
The authors thank Dietrich Krebs and Jan Malte Slowik for helpful discussions.


\section*{References}

\providecommand{\newblock}{}


\end{document}